\documentclass[aps,prl]{revtex4}
\input epsf

\begin{document}

\title{Symmetric Density Functionals}

\author{B.G. Giraud}
\affiliation{giraud@dsm-mail.saclay.cea.fr, Service de Physique Th\'eorique, 
DSM, CE Saclay, F-91191 Gif/Yvette, France}

\date{\today}

\begin{abstract}

Variations in distinct restricted spaces of wave functions generate distinct
density functionals (DF)'s. In particular, angular momentum projected Slater 
determinants define a new DF, compatible simultaneously with angular momentum 
zero and a mean field description. 

\end{abstract}

\maketitle

\bigskip

Consider a finite number of identical particles, fermions for instance. Let
$a^{\dagger}_r$ and $a_r$ be their creation and annihilation operators at 
position $r.$ Their number operator reads, 
${\bf A}=\int dr\, a^{\dagger}_r\, a_r.$ Consider now a nuclear or atomic 
Hamiltonian, ${\bf H} = {\bf T} + {\bf V} + {\bf U},$ 
where $A,$ ${\bf T}=\sum_{i=1}^A t_i,$ ${\bf V}=\sum_{i>j=1}^A v_{ij}$ and 
${\bf U}=\sum_{i=1}^A u_i$ are the particle number, kinetic energy, 
two-body interaction energy and one-body potential energy, respectively. 
It is understood in the following that $u$ is local, 
$\langle r | u | r' \rangle = u(r)\, \delta(r-r'),$ although generalizations 
of the theory to non local potentials are available \cite{Gil}. It is also 
understood that $u$ is a scalar, like $t$ and $v,$ although non scalar $u$'s
will be reinstated at the end of this note.

There is some frustration expressed in the litterature \cite{Dug} at a 
contradiction between the existence theorem provided by Hohenberg and Kohn 
(HK) \cite{HK}, which deals with exact ground states having good quantum 
numbers, and the results given by those functionals used in practice, which 
often provide solutions showing spontaneously broken symmetries. The 
difficulty with such practical functionals comes, obviously, from their close 
connection with mean field theories, such as, for instance the Hartee-Fock 
(HF) variational principle. Moreover, the Kohn-Sham (KS) theorem \cite{KS}, 
uses, orthogonal orbitals of independent particles, although its driving mean 
field does include, in principle, all the influence of correlations in a 
complicated wave function.

That mean field approximations lead to broken symmetries is actually a very 
good feature; there is no doubt indeed, for instance, that Ne$^{20}$ is a 
strongly deformed nucleus! But the subsequent projection of a good angular 
momentum is mandatory; indeed the ground state of Ne$^{20}$ is a $0^+$ and 
the relevant density is spherical, not prolate.

This note proves an existence theorem for a functional that allows 
simultaneously for mean fields breaking symmetries and for densities 
reflecting symmetry restoration. For the sake of generality, we can 
follow Mermin \cite{Mer} and use a finite temperature formalism. 
In Fock space, with a grand canonical ensemble, this means using the second 
quantized forms of ${\bf A}$ and ${\bf H},$ naturally. We first consider 
the domain ${\cal D}$ of all density matrices ${\bf D}$ in Fock space, under 
the obvious normalization constraint of a unit trace. Any approximation such 
as Hartree-Bogoliubov, BCS, shell model mixtures, generator coordinate 
mixtures, finite temperature HF, etc. then 
makes a restriction ${\cal R}$ of the domain ${\cal D}.$ 

The usual variational principle, with a 
temperature $T$ and a chemical potential $\mu,$ reads,
\begin{equation}
F_M={\rm Min}_{{\bf D} \in {\cal D}}\,F,\ \ \ \ \ 
F = \frac{ Tr\, {\bf D}\,  
\left( {\bf H} - \mu\, {\bf A} + T\, \ln {\bf D} \right) } { Tr\, {\bf D} }\, .
\end{equation}
With unrestricted density operators, it selects the well known, unique 
position, 
${\bf D}_M=\exp \left[ - ({\bf H} - \mu\, {\bf A} )/T \right]/ Z,$ with 
$Z= Tr\, \exp \left[ - ({\bf H} - \mu\, {\bf A} )/T \right],$ 
where a smooth minimum, $F_M,$ is reached. Any variation $\delta {\bf D}$ 
away from ${\bf D}_M$ creates a strictly positive increase of $F$ at 
second order with respect to $\delta {\bf D}.$ The functional double
derivative, $\delta^2 F/(\delta {\bf D}\, \delta {\bf D}'),$ is 
strictly positive definite at that position ${\bf D}_M$ in ${\cal D}.$ 
Accordingly, in any, whether local or global, system of coordinates for 
${\cal D},$ the matrix representing 
$\delta^2 F/(\delta {\bf D}\, \delta {\bf D}')$ will be invertible.

From the very definition of ${\bf D}_M,$ the first order functional 
derivative, $\delta F/\delta {\bf D},$ vanishes at that position. Hence, if
one introduces an infinitesimal variation $\delta u,$ triggering an 
infinitesimal displacement of ${\bf D}_M,$ the only contribution to 
$\delta F_M$ comes from $\delta u.$ In short, 
$\delta F_M = Tr\, {\bf D}_M\ \delta {\bf U}.$ There is no contribution from 
$\delta {\bf D}_M.$ 

Define the one-body density matrix in coordinate representation,
$\nu(r,r')=Tr\, {\bf D_M}\, a^{\dagger}_r\, a_{r'}.$
Its diagonal, $\rho(r)=\nu(r,r),$ is the usual density deduced from 
${\bf D}_M$ by integrating out all particles but one. Hence 
$\int dr\, \rho(r) = Tr\, {\bf D}_M\, {\bf A}.$ Note also the result, 
$\delta F_M = \int dr\, \rho(r)\, \delta u(r).$ In other words, 
$\delta F_M/\delta u = \rho.$

Freeze $t$ and $v$ and consider $F_M$ as a functional of $u$ alone. The 
Hohenberg-Kohn-Mermin (HKM) process then consists in a Legendre tranform of 
$F_M,$ based upon the essential result,
\begin{equation}
\frac{\delta F_M}{\delta u}=\rho.
\label{funcderi}
\end{equation}
This Legendre transform involves two steps, namely i) subtract from $F_M$
the functional product of $u$ and $\delta F_M/\delta u,$ namely the integral, 
$\int dr\, u(r)\, \rho(r),$ leaving 
\begin{equation}
{\cal F}_M = Tr\, {\bf D_M}\, 
\left( {\bf T}+{\bf V}-\mu\, {\bf A}+T\, \ln {\bf D_M} \right),
\label{exctgs}
\end{equation}
then ii) consider $\rho$ as the primary ``variable'' rather than $u,$ namely 
consider ${\cal F}_M$ as a functional of $\rho.$ Step ii) is made possible 
by the one-to-one correspondence between $u$ and $\rho,$ under precautions 
such as the exclusion of those trivial variations $\delta u$ which modify 
$u$ by a constant only. (For some discussion of more precautions, see for 
instance \cite{Leeu} and \cite{Gir}.) The one-to-one correspondence is proven 
by an argument {\it ab absurdo} : if distinct potentials $u$ and $u'$ 
generated (distinct!) ${\bf D}_M$ and ${\bf D}_M^{\, \prime}$ with the same 
$\rho,$ then two contradictory, strict inequalities would occur, namely,
\begin{equation}
\int dr\, [\, u(r)-u'(r)\, ]\, \rho(r) < F_M - F_M^{\, \prime}\ \ \ 
{\rm and}\ \ \  
\int dr\, [\, u(r)-u'(r)\, ]\, \rho(r) > F_M - F_M^{\, \prime}.
\label{absurd}
\end{equation} 

An inverse Legendre transform, returning from ${\cal F}_M$ to $F_M,$ is then 
available, according to the property,
\begin{equation}
\frac{\delta {\cal F}_M}{\delta \rho}=-u.
\end{equation}

Now, the {\it same} properties and process hold if we assume that, when 
${\bf D}$ is restricted to a smaller domain ${\cal R},$ there is still a 
{\it unique} position ${\bf D_m}$ for a {\it smooth again}, strict minimum, 
$F_m={\rm Min}_{{\bf D} \in {\cal R}}\, F.$ The subscript $M$ should be 
replaced by the subscript $m,$ but otherwise, it is trivial to recover
Eq. (\ref{funcderi}) and the contradiction (\ref{absurd}) to implement
a one-to-one correspondence and a Legendre transform.

This will be illustrated by the special case of angular momentum restoration,
at zero temperature. Limits at $T=0$ are easy and do not need much 
discussion. Consider therefore a canonical situation, where is no need
for $\mu;$ there is a well defined particle number $A,$ and density operators 
are dyadics, ${\bf D}=| \psi \rangle \langle \psi |,$ where $| \psi \rangle$ 
is just an $A$-particle wave function. The variational principle, Eq. (1), 
trivially reduces to the Rayleigh-Ritz form,
\begin{equation}
F_M={\rm Min}_{\psi}\, F\, ,\ \ \ \ \ F = 
\frac{ \langle \psi | {\bf H} | \psi \rangle } { \langle \psi | \psi \rangle }
\, ,
\end{equation}
and the minimum with unrestricted wave functions generates the exact ground 
state with its exact eigenvalue $F_M,$ parametrized by $u.$
Restrict now the wave functions to be angular momentum projected Slater 
determinants, 
\begin{equation}
\Psi_0=\int d\Omega\, {\bf R}(\Omega)\, | \phi \rangle,
\label{restrict}
\end{equation}
where $\Omega$ and ${\bf R}$ label rotation angles and rotation  
operators, respectively, and $| \phi \rangle$ is an 
unrestricted, hence most often deformed, Slater determinant. Angular momentum 
$0$ has been chosen here because every even-even nucleus has 
such a ground state. In a short, transparent  notation, 
$| \Psi_0 \rangle = {\bf P}_0\, |\phi \rangle,$ where ${\bf P}_0$ can be 
normalized to be idempotent, ${\bf P}_0^2={\bf P}_0.$ It commutes with 
${\bf H},$  a scalar, and we use these commutation and idempotence in the 
following.

The key question is now whether there is a unique $\Psi_{0m}$ for which the 
minimum, 
\begin{equation}
F_m = {\rm Min}_{\Psi_0}\, 
\frac{ \langle \Psi_0 | {\bf H} | \Psi_0 \rangle } 
     { \langle \Psi_0 |           \Psi_0 \rangle }\, ,
\end{equation}
is reached. Notice that, in any case, the restriction of trial functions 
enforces an inequality, $F_m \ge F_M,$ and that this inequality is generally 
strict, $F_m > F_M\, ;$ it demands very special Hamiltonians and potentials 
for an equality to occur.

For deformed nuclei the same minimum $F_m,$ which also reads,
\begin{equation}
F_m = {\rm Min}_{\phi}\, 
\frac{ \langle \phi | {\bf H}\ {\bf P}_0 | \phi \rangle } 
     { \langle \phi | {\bf P}_0                          | \phi \rangle }
\, ,
\label{projvar}
\end{equation}
occurs for many choices of $\phi\, .$ All such Slater determinants are 
deduced from one another by rotations. Hence they generate the same 
projected state $\Psi_0\, .$ As illustrated by Figure 1, that breach of 
uniqueness, namely the rotational degeneracy, caused at the $\phi$ level by 
deformation, is completely corrected at the $\Psi_0$ level. There might be 
further, accidental causes of degeneracies of the minimum in the $\Psi_0$ 
space, but it is clear that the broken symmetry degeneracy has been removed 
by the projector ${\bf P}_0\, .$

It is thus reasonable to assume that, provided all those symmetries broken by 
HF and similar approximations have been restored, a search for a minimum 
energy, {\it after} the restoration, will suffer from no degeneracy any more.

\begin{figure}[htb] \centering
\mbox{  \epsfysize=100mm
         \epsffile{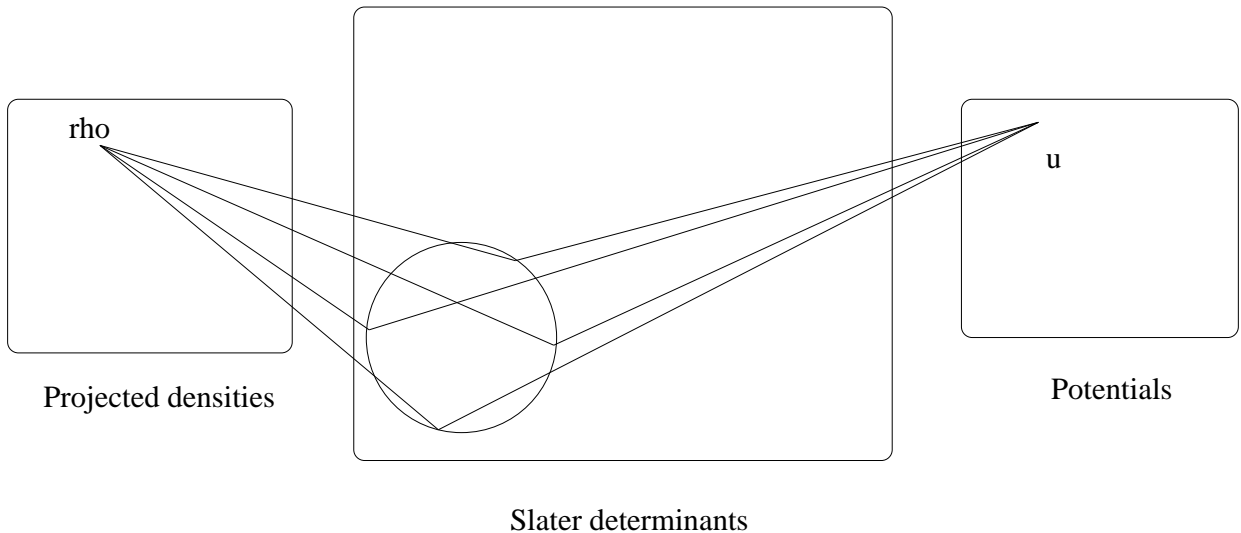}
     }
\caption{One-to-one correspondence $u \leftrightarrow \rho.$ A given potential 
$u$ (right) induces degenerate Slater determinant variational solutions 
(center) resulting in the same density $\rho$ after angular momentum 
projection (left).}
\end{figure}

Under this assumption of uniqueness of the minimum, consider the space of
those $\Psi_0$ that are square normalized to unity. (This technicality spares
us denominators.) Let $\Psi_{0m}$ and $\Psi_{0m}^{\, \prime}$ be those
states which provide $F_m$ and $F_m^{\, \prime}$ for ${\bf H}$ and ${\bf H}'$
driven by distinct $u$ and $u',$ respectively. Then, the same argument {\it
ab absurdo} as that for (\ref{absurd}), with the same precaution of a non 
trivial difference between $u$ and $u',$
\begin{eqnarray}
\int dr\, [\, u(r)-u'(r)\, ]\ \rho(r) &=& 
\langle \Psi_{0m} | \left( {\bf H} - {\bf H}' \right) | \Psi_{0m} \rangle 
< F_m - F_m^{\, \prime}\, , 
\nonumber \\  
\int dr\, [\, u(r)-u'(r)\, ]\ \rho(r) &=&
\langle \Psi_{0m}^{\, \prime} | \left( {\bf H} - {\bf H}' \right) | 
\Psi_{0m}^{\, \prime} \rangle  
> F_m - F_m^{\, \prime}\, .
\label{absurdagain}
\end{eqnarray} 
proves that $\Psi_{0m}$ and $\Psi_{0m}^{\, \prime}$ cannot induce the same
$\rho.$ Hence the one-to-one map, $u \leftrightarrow \rho,$ holds 
for the restricted variation space. It must be stressed here that the 
density $\rho$ is that of the symmetry projected state. It can differ from the
density of every parent state $\phi$ through the correlations brought by 
the coherent summation over angles, Eq. (\ref{restrict}).

The ``smoothness'' condition of $F$ around its minimum is necessary to allow 
for functional derivatives $\delta F_m/\delta u,$ 
$\delta^2 F_m/(\delta u\, \delta u'),$ etc. Since there are are no 
singularities in a  space  of Slater determinants (although this space is 
curved), and since that subset of determinants which might create 
singularities because of a vanishing matrix element 
$\langle \phi | {\bf P}_0 | \phi \rangle$ has obviously a zero measure, we 
recover all the properties of the unrestricted variational space. In 
particular, $\delta F_m/\delta u=\rho.$ A Legendre transform is possible. 
Hence there exists a functional of a {\it spherical} $\rho,$ with value
\begin{equation}
{\cal F}_m = \langle \Psi_{0m} | ( {\bf T} + {\bf V} ) | \Psi_{0m} \rangle.
\label{apprgs}
\end{equation}
It may be stressed here how a spherically invariant $u$ is in a one-to-one
correspondence with a spherically invariant $\rho.$ This is reinsuring for the 
consistency of this existence theorem, deduced from a restricted variational 
space. 

Since $\Psi_{0m}$ can only be an approximate ground state, then
${\cal F}_m > {\cal F}_M,$ compare Eq. (\ref{exctgs}) without $\mu$ at $T=0$ 
and Eq. (\ref{apprgs}). We are thus dealing with a new functional. It can only 
give upper bounds to bound state energies, but this is expected for 
restricted variational spaces, obviously.

For the case of nuclear forces, with their very repulsive core and need for 
short range correlations, nothing prevents us, formally at least, from 
refining the variational space by introducing a ``correlator operator'' 
{\it a la} Jastrow for instance, that crushes the wave function at short 
relative distances between particles,
\begin{equation}
F_m = {\rm Min}_{\phi}\, 
\frac{ \langle \phi | {\bf C}\, {\bf H}\, {\bf C}\ {\bf P}_0 | \phi \rangle } 
     { \langle \phi | {\bf C}^2\                   {\bf P}_0 | \phi \rangle }
\, ,\ \ \ \ \ 
{\bf C}= \prod_{i>j=1}^A \chi\left( |r_i-r_j| \right),
\end{equation}
where $\chi$ vanishes at the origin and becomes $1$ at large distances. Since
${\bf C}$ is a scalar, it commutes with ${\bf P}_0.$

Our considerations can be easily generalized to finite temperatures, and
to other symmetry restorations than angular momentum projections, and to other 
restricted spaces. It can be concluded that every restricted space and every 
symmetry restoration creates its own density functional, provided the 
conditions of uniqueness and smoothness of the variational minimum are 
fulfilled. For the discussion of differentiability and fine topological 
properties of the $u$- and $\rho$ spaces we refer again to \cite{Leeu}.
Up to our understanding, the validity of our existence theorems for symmetry 
conserving functionals is not compromised.

Such existence theorems, though, suffer from the usual plague of 
the field: constructive algorithms are missing.

\bigskip
Another question must now be raised: can one define a KS theory compatible 
with symmetry restoration? Consider again the case of angular momentum $0,$ 
for non degenerate ground states of scalar Hamiltonians, with $\rho$ scalar, 
as are $t,$ $v$ and $u.$ One may then claim the exact $F_M$ and its partner 
${\cal F}_M$ are also scalar. Accordingly, all the tools of the KS theory, 
including the ``under-rug-sweeping'' exchange-correlation potential 
$v_{xc}(r),$ must be rotationally invariant. For the sake of simplicity, let 
us forget spin-orbit complications. Then the KS equations read in coordinate 
space only,
\begin{equation}
\left[ t + \int dr'\, v(r-r')\, \rho(r') + v_{xc}(r) + u(r) - 
\varepsilon_{\ell} \right]\, \varphi_{\ell m}(r) =0,
\label{KSeqsym}
\end{equation}
Can they be assumed to generate the same density as that of the correlated 
ground state?

In Eqs. (\ref{KSeqsym}), the magnetic degeneracy of the orbital eigenvalues
$\varepsilon_{\ell}$ is explicit. A complete filling of a core, made of all 
lowest shells but one, and a homogeneous, partial filling of a cloud for the 
residual particle number $z$ in the next, partly filled shell, seem to be in 
order. It is trivial indeed that the following density,
\begin{equation}
\rho(r)=\sum_{i \in core} |\varphi_i(r)|^2\ + \frac{z}{2L+1}\,
\sum_{M \in cloud} |\varphi_{LM}(r)|^2,
\label{fill}
\end{equation}
is rotationally invariant. Here $L$ is the orbital label for the cloud, 
that first shell above the filled core, and $M$ labels the corresponding 
degenerate orbitals. Can such a density be related to the angular momentum 
projection of any state in the subspace spanned by all the 
%$\left(\matrix{2L+1 \cr z}\right)$ 
degenerate Slater determinants available 
when filling $z$ among the degenerate $2L+1$ orbitals? Notice, incidentally, 
that necessarily $z \ne 1$ and $z \ne 2L,$ since such situations would only 
allow total angular momentum $L.$ Notice also that particle-hole symmetry that 
relates the coupling of angular momenta in a situation with 2L+1-$z$ to that 
with $z.$ For simplicity, we thus set $2 \le z \le L.$ Then we forget about
the core and investigate the densities of those angular momentum projected 
determinants made of mixed orbitals inside the cloud. By mixed orbitals, we
mean arbitrary linear superpositions of the magnetic numbers $M;$ such 
mixtures play the r\^ole of deformations. What is the density of such projected
``deformed'' determinants?

Consider $L=6,\, z=4\, $ for an illustrative example. Standard angular 
momentum techniques prove that one can then make two distinct states, 
$\Psi_{0a}$ and $\Psi_{0b},$ with total angular momentum $0.$ Therefore the 
projected Slater determinants must be linear combinations 
$\alpha\, \Psi_{0a} + \beta\, \Psi_{0b},$ with densities $\alpha^2\, \rho_a +
\beta^2\, \rho_b + 2\, \alpha\, \beta\, \rho_{ab},$ where $\rho_{ab}$ is the 
``transition density'',
\begin{equation}
\rho_{ab}(r) = 
\langle\, \Psi_{0b}\, |\, a^{\dagger}_r\, a_r\, |\, \Psi_{0a}\, \rangle,
\end{equation}
while $\rho_a$ and $\rho_b$ are the densities of $\Psi_a$ and $\Psi_b,$ 
respectively. The latter densities are the same, however. Indeed, since all
orbitals have the same radial shape, $\varphi\left(|r|\right),$ there are
only angular integrals to consider when calculating the densities. For
instance,
\begin{equation}
\rho_a(r) = 4\, \varphi(|r|)^2 \int d\hat r_2\, d\hat r_3\,  d\hat r_4\,
| \psi_a\left( \hat r, \hat r_2, \hat r_3,  \hat r_4 \right) |^2,
\end{equation}
where $\hat r_i$ designates the polar angles of a particle $,$ and $\psi_a$ 
carries the angular properties of $\Psi_{0a}.$ Because of the scalar nature of 
$\rho_a\, ,$ it is obvious that the angular integral gives a result, 
$\rho_a(r)=4\, \varphi(|r|)^2,$ which does not depend on the singled out 
$\hat r.$ The same holds for $\Psi_{0b}.$ Moreover, the transition density 
reads,
\begin{equation}
\rho_{ab}(r) = 4\, \varphi(|r|)^2 \int d\hat r_2\, d\hat r_3\,  d\hat r_4\
 \psi_b\left( \hat r, \hat r_2, \hat r_3,  \hat r_4 \right)^*\ \,
 \psi_a\left( \hat r, \hat r_2, \hat r_3,  \hat r_4 \right),
\end{equation}
and this must again reduce to a constant with respect to $\hat r$ because
of the rotation invariance of both $\Psi_{0a}$ and  $\Psi_{0b}.$ The constant,
however, must vanish. This is because, upon an integration with respect to
$r,$ one must recover the orthogonality of $\Psi_{0a}$ and  $\Psi_{0b}.$
It can be concluded that every ``deformed'' determinant made of orbitals
taken from the spherical ``cloud'' shell has the same density. In 
Eq. (\ref{fill}), the sometimes called ``filling'' approximation term,
\begin{equation}
\frac{z}{2L+1}\, \sum_{M \in cloud} |\varphi_{LM}(r)|^2 = 
z\ \varphi\left(|r|\right)^2,
\end{equation}
is not an approximation; it actually makes the exact ansatz for an exact,
projected density.

The burden of an accurate description of $\rho$ then falls upon the radial
shape $\varphi(|r|),$ that is the output of the KS equations, 
Eqs. (\ref{KSeqsym}). This makes it all the more important to better describe
the often badly known potential $v_{xc}.$

Generalizations to other values of $L$ and $z$ and to more than two independent
states with total angular momentum zero are obvious. All determinants from the 
cloud shell give the same density after projection.

To conclude, a word is in order about non spherical potentials $u,$ and more 
generally symmetry breaking potentials $u.$ As long as one is interested in 
Eq. (\ref{projvar}), or rather in that proper form that takes into account the
fact that $u$ does not commute any more with the symmetry restoring projector 
${\bf P},$ (still normalized to be idempotent,)
\begin{equation}
F_m = {\rm Min}_{\phi}\, 
\frac{ \langle \phi |\, {\bf P} {\bf H}\, {\bf P}\, | \phi \rangle } 
     { \langle \phi | {\bf P}                       | \phi \rangle }
\, ,
\end{equation}
it is clear that only that symmetric part, ${\bf P} {\bf U} {\bf P},$ is
activated. Hence a non symmetric $u$ brings nothing. The situation is different
for the KS theory. Indeed, both the Hartree-like field, 
$\int dr'\, v(r-r')\, \rho(r'),$ and a symmetry violating $u$ lift, in general,
the unlikely degeneracies of the KS orbitals, and an unprojected density 
results from,
\begin{equation}
\left[ t + \int dr'\, v(r-r')\, \rho(r') + v_{xc}(r) + u(r) - 
\varepsilon_i \right]\, \varphi_i(r) =0,\ \ \ \ \ 
\rho(r)=\sum_i |\varphi_i(r)|^2.
\label{KSeqnosym}
\end{equation}
The sum, $\sum_i\, ,$ is taken, naturally, upon the lowest orbitals, except in 
very few anomalous cases. Consider a limit where $u$ becomes symmetric. As is 
well known, there is no reason why, at that limit, a symmetric state should 
emerge out of Eqs. (\ref{KSeqnosym}). One needs an ansatz like Eq. (\ref{fill})
to enforce the symmetry. Under such an ansatz, which should be tuned for the 
partial filling of quasi degenerate orbitals, as with a finite temperature,
the Hartree potential and a slightly symmetry violating $u$ will indeed create 
a slight degeneracy lifting only. There is thus a serious difference between 
an unconstrained KS approach and the symmetry constrained one.

To summarize this note, two distinct results are claimed, namely i) existence 
theorems for density functionals explicitly enforcing a symmetry by means of
projectors chiseling variational spaces, and ii) a symmetry respectful
version of the KS theory. The latter comes from an adjustment of the definition
of the density; KS orbitals need to be incoherently mixed to account for 
symmetry projection.

It is a pleasure to thank T. Duguet for a critical reading of this note. B.G.G.
also thanks the hospitaliy of the University of Hokkaido, where part of this 
work was done.

\end{document}